\documentclass[twocolumn]{revtex4}
\usepackage{amsmath}
\usepackage{amssymb}
\usepackage{graphicx}
\usepackage{bm}
\usepackage{dcolumn}
\usepackage[cp1251]{inputenc}

\begin{document}

\title{Inelastic neutron scattering as a confirmation of a new type of gapped surface
excitations in liquid helium}
\author{P. D. Grigoriev$^{1,2}$, A. D. Grigoriev$^{3}$, A. M. Dyugaev$^{1}$}
\affiliation{$^{1}$L.D. Landau Institute for Theoretical Physics, Chernogolovka, Russia}
\affiliation{$^{2}$National University of Science and Technology "MISiS", 119049 Moscow,
	Russia} 
\affiliation{$^{3}$Samara State Technical University, 443100, Samara, Russia}
%\email{grigorev@itp.ac.ru}
\begin{abstract}
We analyze the experimental data \cite{GodfrinPrivate} on inelastic neutron scattering by a thin ~5-atomic-layer film of liquid helium at three different temperatures: T=0.4K, 0.98K and 1.3K. These data were partially published previously,\cite{Godfrin1992,GodfrinJLTP1992,Clements1996} but here we present them in a better quality and at various temperatures. The neutron scattering intensity plots, in addition to the previously know dispersion of phonons and ripplons, suggest a branch of gapped surface excitations with activation energy $\sim 4.5$K and the dispersion similar to that expected for surfons -- the bound quantum states of helium atoms above liquid helium surface, proposed and investigated theoretically \cite{SurStates,SurStates2011}. These data, probably, provide the first direct experimental confirmation of surfons. Before these surface excitations received only indirect experimental substantiation, based on the temperature dependence of surface tension coefficient \cite{SurStates,SurStates2011} and on their interaction with surface electrons \cite{WeJETPLett2008,MobilityLett}. The existence of surfons as an additional type of surface excitations,  although being debated yet, is very important for various physical properties of He surface. We also analyze previous numerical results on excitations in liquid helium and argue that surface excitations similar to surfons have been previously obtained by numerical calculations and called resonance interface states \cite{Gernoth1992}. 
\end{abstract}
\date{\today}

\maketitle

%\author{E. Krotscheck}
%\affiliation{Department of Physics, Texas A\& M University, College Station, Texas 77843}

\section{Introduction}

A deep understanding of the processes on the surface of liquids is important
for various fields of natural science: physics, chemistry, biology. The
microscopic description of liquid surface is a rather complicated
problem, and various theoretical techniques have been applied to advance it.
\cite{SurfTensionBook} At low temperature the quantum nature of surface
excitations is important, which becomes apparent in liquid helium and can be
experimentally studied, e.g., using the interaction of these excitations
with surface electrons \cite{EdwardsReview1978,Edelman,Shikin,Monarkha}. In
bulk liquid helium the excitations are well known both from microscopic
theory \cite{FenbergBook1969,Mahan} and from the extensive thermodynamic and
neutron scattering experiments. The microscopic description of the surface
excitations in liquid helium is more complicated because of spatial
inhomogeneity of this problem. This problem was quite successfully studied
using the numerical variation methods with the Feenberg wave function in the
so-called hypernetted-chain approximation \cite%
{Krotscheck1985II,Krotscheck1985Substrate,KrotscheckTymczak,ClementsKrotscheckPRB1996,Gernoth1992,Gernoth1994}%
. These numerical results were used to analyze the experimental data on inelastic neutron scattering by liquid
helium films,\cite{Lauter,Godfrin1992,GodfrinJLTP1992,Clements1996,ClementsKrotscheckJLTP1997} the temperature dependence of the surface tension coefficient \cite{Pricaupenko} and other thermodynamic properties \cite{CampbellPRB1997}.

In liquid $^{4}$He there is only one type of gapless surface excitations --
the quanta of surface waves, called ripplons. For the wavelength much larger
than interatomic distance but smaller than the capillary length $\kappa
^{-1}\approx 0.05$cm, the dispersion of surface waves is given by \cite{LL6} 
\begin{equation}
\omega _{q}^{2}=\frac{\alpha }{\rho }q^{3}~\tanh \left( qd\right) ,
\label{2}
\end{equation}%
where $\alpha $ is surface tension coefficient, $\rho $ is the density of
liquid, $q$ is the ripplon wave number, and $d$ is the width of liquid
helium film. For short-wave-length ripplons with $q\gtrsim 1$\AA $^{-1}$ the
ripplon dispersion $\omega (q)$ becomes softer than in Eq. (\ref{2}) and saturates at energy $%
\hbar \omega _{D}\approx 0.8meV\approx 10K$, which was obtained numerically
\cite%
{Krotscheck1985II,Krotscheck1985Substrate,KrotscheckTymczak,ClementsKrotscheckPRB1996,Gernoth1992,Gernoth1994}
and observed using the inelastic neutron scattering by liquid helium films.
\cite{Godfrin1992,GodfrinJLTP1992,Clements1996}

Recently, an additional new type of surface excitations has been proposed
semi-phenomenologically to explain too strong temperature dependence of
surface tension coefficient $\alpha \left( T\right) $.\cite%
{SurStates,SurStates2011} These excitations, called \textit{surfons}, can be
considered as the quantum states of He atoms localized above liquid surface.
Surfons resemble the Andreev states of $^{3}$He atoms in the $^{3}$He-$^{4}$%
He mixture \cite{AndreevLevels}, or the states of $^{3}$He and $^{4}$He atoms
on the surface of liquid hydrogen.\cite{Hydrogen1} According to this
phenomenological model \cite{SurStates,SurStates2011}, the surfons are
localized only along the $z$-axis perpendicular to helium surface and can propagate
along the surface. Their dispersion is 
\begin{equation}
\varepsilon (k)\approx \Delta +k_{\parallel }^{2}/2M^{\ast },  \label{Disp}
\end{equation}%
where $k_{\parallel }$ is a 2D surfon momentum along the surface, $\Delta
> 0$ is the activation energy of a surfon at $k_{\parallel }=0$, and
their effective mass $M^{\ast }$ is of the order of the atomic He mass $%
M_{4}^{0}=6.7\cdot 10^{-24}$g. The surfon activation energy $\Delta $ is
weakly temperature-dependent, 
\begin{equation}
\Delta \left( T\right) =E_{s}-\mu \left( T\right) ,  \label{Delta}
\end{equation}%
where $\mu \left( T\right) $ is the temperature-dependent chemical potential
of a liquid $^{4}$He, $\mu \left( T=0\right) \approx -7.17K$, and $E_{s}$\
is the discrete energy level of He atom at liquid surface, dressed by the
interaction with other atoms, ripplons and phonons. At low temperature $T\ll
\Delta $ the surfon concentration is exponentially small.

Currently there are several experimental facts which can be treated as indirect 
substantiation of surfons. The first two are related to the interaction of
surface electrons with surfons, which provides an additional
temperature-dependent scattering mechanism of surface electrons. This
additional scattering mechanism can considerably improve \cite{MobilityLett}
the agreement between the observed \cite{ExpMobil} and calculated \cite{Saitoh} mobility
of surface electrons. Surfons can also explain the observed \cite{Edelman}
temperature-dependent shift of the transition line between two lowest
electron states above liquid helium or solid hydrogen surface.\cite{WeJETPLett2008} 
Finally, the surfons can
explain \cite{SurStates,SurStates2011} the long-standing puzzle of the 
observed \cite{HeDataBook} too strong temperature dependence $%
\alpha \left( T\right) $ of the surface tension coefficient of liquid He. 
The comparison can be used to estimate the surfon
activation energy for both He isotopes from experiment as \cite{SurStates2011}%
\begin{equation}
\Delta ^{He4}\approx 2.67K,~\Delta ^{He3}\approx 0.25K  \label{D}
\end{equation}%
corresponding to $E_{s}^{He4}\approx -4.5K$ and $E_{s}^{He3}\approx -2.25K$.
This value $\Delta ^{He4}\approx 2.67K$, obtained from the fitting of $%
\alpha \left( T\right) $,\cite{SurStates2011} is in a reasonable agreement
with the energy gap $\Delta _{2}\approx 1.6K$ of resonance interface states
obtained from the numerical calculations in Ref. \cite{Gernoth1992} (see
below) and with $\Delta ^{He4}\approx 3.2K$ obtained from the
semi-phenomenological description in Ref. \cite{SurStates2011}. Fitting the
surface tension of a thick He film, assuming that only ripplons and surfons
make considerable contribution, gives an upper estimate of the surfon
effective mass:\cite{SurStates2011} $M_{4}^{\ast }\approx 2.65M_{4}^{0}$,\
and $M_{3}^{\ast }\approx 2.25M_{3}^{0}$, where $M_{4}^{0}=6.7\cdot 10^{-24}$%
g and $M_{3}^{0}=5.05\cdot 10^{-24}$g are the free atomic masses of $^{4}$He
and $^{3}$He correspondingly. The effective surfon mass $M^{\ast }$ in its  
in-plane motion is, probably, renormalized by interaction with liquid.
Note, that the inclusion of the second branch of surface excitations,
obtained numerically in Ref. \cite{Gernoth1994}, also
considerably improves the agreement between experiment \cite{HeDataBook} and
theory \cite{Pricaupenko} on $\alpha \left( T\right) $ of thick He films. The
temperature dependence of surface tension coefficient $\alpha \left(
T\right) $ of thin He films cannot also be fitted without additional type of
surface excitations, though the \textquotedblright breathing
modes\textquotedblright\ (or \textquotedblright quantized bulk
phonons\textquotedblright ), obtained in Refs. \cite%
{Krotscheck1985II,Krotscheck1985Substrate,KrotscheckTymczak,ClementsKrotscheckJLTP1997}
make a considerable contribution \cite{KrotscheckPrivate} to $\alpha \left(
T\right) $.

The microscopic substantiation of the surfon existence was provided 
\cite{SurStates2011} by the solution of one-particle Schr\"{o}dinger equation for
a He atom above liquid He surface in the effective one-dimensional potential 
$V\left( z\right) $ created by the interaction with other He atoms in the lower
half-space (Hartree approximation). The corresponding one-particle Schr\"{o}%
dinger equation is uniform in the $x$-$y$ plane and does not take into account the correlation
effects. However, it definitively gives a discrete quasi-stationary energy level $%
E_{s}\approx -1.24K<0$ of a He atom above the surface. The neglected correlation
effects in liquid He, which partially may be treated as a
\textquotedblright dressing\textquotedblright\ of a surfon by ripplonic
polaron, reduce considerably the value $E_{s}$ of this quasi-stationary
energy level to $E_{s}\approx -4K$, as was semi-phenomenologically estimated
in Ref. \cite{SurStates2011}, but they do not destroy these excitations.
This quasi-stationary level also persists after the inclusion of
exchange interaction between He atoms. The latter is weak for liquid $^{4}$He
because the wave functions of He atoms overlap weakly due to their
strong hard-core repulsion at distance $z<2.5\mathring{A}$.

The surfon lifetime $\tau $ is rather short and limited mainly by two
processes: the immersion into liquid and evaporation due to scattering by
other excitations. The second process was studied in Ref. \cite%
{SurfonEvaporation}. The evaporation rate $1/\tau _{v}$ of surfons depends
on their initial momentum along the surface and grows rapidly with the
increase of temperature.\cite{SurfonEvaporation} However, below $4$K it does
not exceed the limit $\sim $ $E_{s}/\hbar $ where the surfons cannot be
called quasiparticles. The immersion rate of surfons to the liquid has not
been calculated yet, but it should also be less than $E_{s}/\hbar $, because
in order to sink into the liquid, a He atom must overcome a potential barrier and
rearrange the surrounding atoms of the liquid.

Thus, the surfons are, presumably, non-stable quasi-particle with lifetime
shorter than that of long-wave-length ripplons or phonons. Nevertheless, the
existence of surfons as an additional type of surface excitations, 
although being debated yet, is crucial for various physical 
properties of He surface. In addition to
explaining the strong temperature dependence of the surface tension coefficient,%
\cite{SurStates,SurStates2011} they may considerably increase the
evaporation rate of liquid He by adding a new evaporation channel via the
intermediate surfon state with activation energy $\Delta $ smaller than the
evaporation energy $\left\vert \mu \right\vert $.\cite{SurfonEvaporation} 
The surfon quasi-stationary quantum states may also affect the reflection coefficient of He atoms by liquid He surface \cite{TuckerJLTP1995}. 
Therefore, any substantiation of this new type of excitations, experimental
or theoretical, is very important. At the moment there are only indirect
experimental confirmations of the existence of surfons by their interaction
with surface electrons \cite{WeJETPLett2008,MobilityLett} or their
contribution to surface tension coefficient \cite{SurStates,SurStates2011}. 
In this paper we analyze the experimental data on inelastic
neutron scattering by thin He films and study if these data can provide a
direct experimental confirmation of the existence of surfons. We also
summarize the available results of ab-initio numerical calculations of surface
excitations in liquid He, which also indicate the existence of surfons.

\begin{figure*}[tbh]
\includegraphics[width=0.99\textwidth]{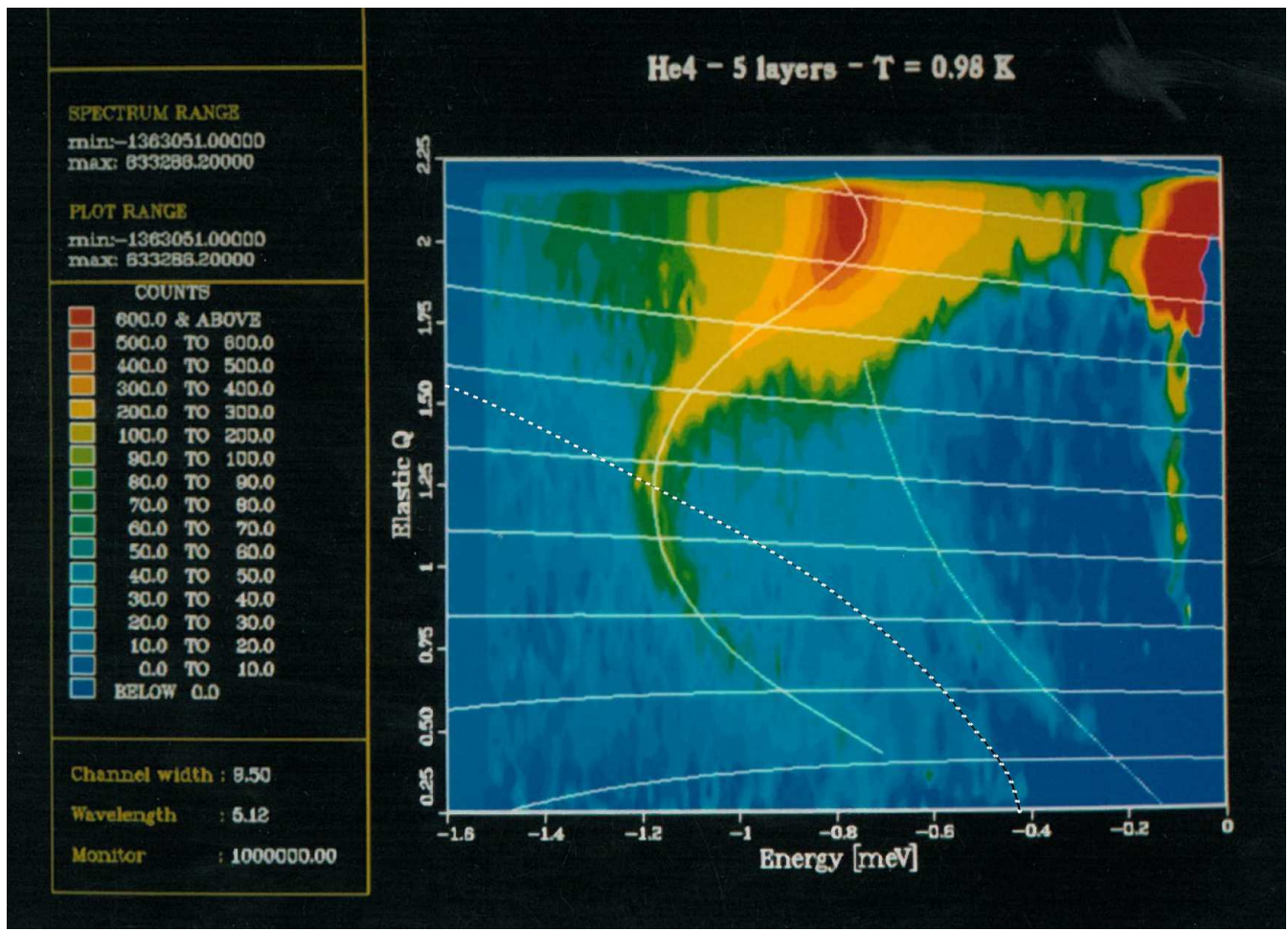}
\caption{(Color on-line) Experimental data on neutron inelastic scattering
intensity by liquid helium film, containing only 5 atomic layers at $T=0.98K,
$ as a function of the in-plane momentum $Q=q_{\parallel }$ and of the energy
transfer. The solid white lines along the intensity maxima mark the phonon and ripplon spectra. 
The white dashed line corresponds to the expected dispersion of surfons.}
\label{FigT098}
\end{figure*}
%\end{widetext}

\section{Experimental data and their analysis}

%(see Fig. \ref{SurfTensFilm})
%\begin{figure}[tbh]
%\includegraphics[width=0.49\textwidth]{sigma.eps}
%\caption{The comparison with experiment of the theoretical predictions of the temperature dependence of surface tension coefficient $\protect\alpha %\left( T\right) $ of thin He films, which takes into account ripplons and quantized bulk phonons obtained in
%Ref. \protect\cite{ClementsKrotscheckJLTP1997}.}
%\label{SurfTensFilm} \end{figure}

\begin{figure}[tbh]
\includegraphics[width=0.49\textwidth]{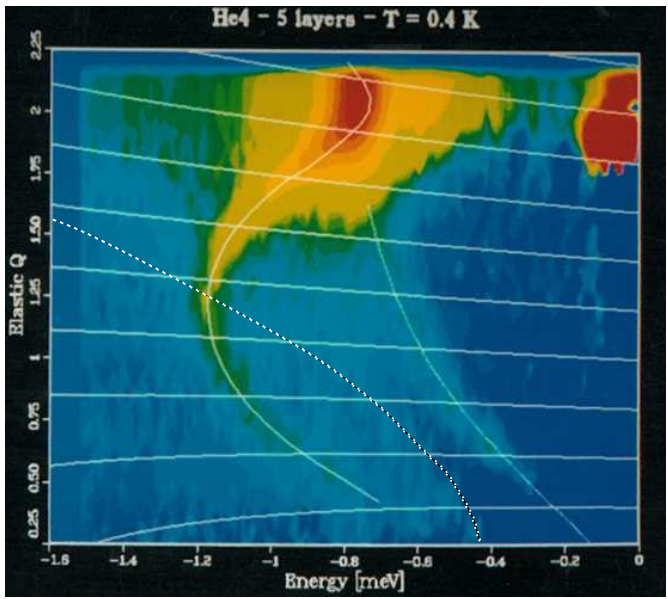}
\caption{(Color on-line) Experimental data on neutron inelastic scattering
intensity by liquid helium film, containing only 5 atomic layers at $T=0.4K,$
as function of the in-plane momentum $Q=q_{\parallel }$ and of energy
transfer. The white dashed line corresponds to the expected dispersion of
surfons.}
\label{FigT04}
\end{figure}

\begin{figure}[tbh]
\includegraphics[width=0.49\textwidth]{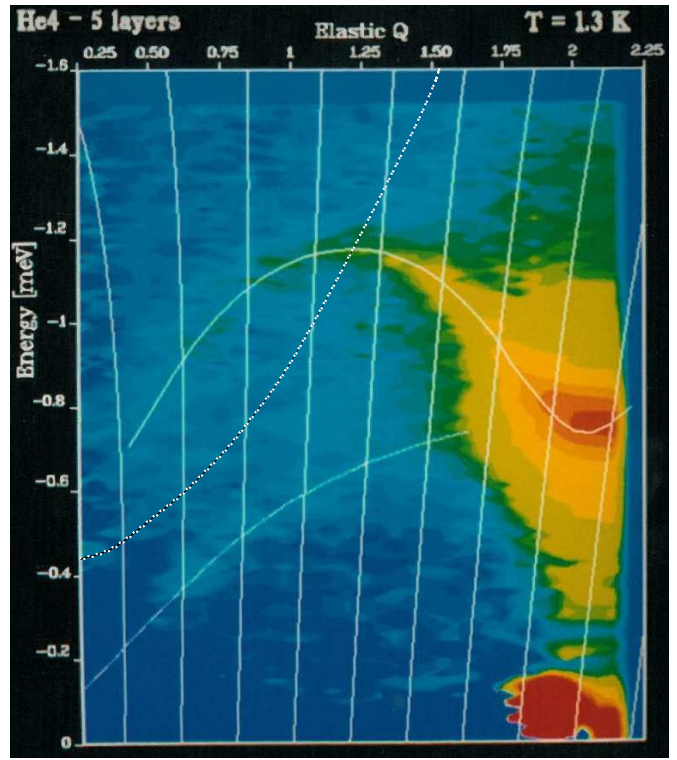}
\caption{(Color on-line) The same as in Figs. \ref{FigT098}-\ref{FigT04} but at different temperature $T=1.3K$ and rotated by $90^{\circ}$.}
\label{FigT13}
\end{figure}

In this section we present and analyze the experimental data, obtained by the group of H. Godfrin \cite{GodfrinPrivate} and shown in Figs. \ref{FigT098}- \ref{FigT13},  on inelastic neutron scattering by a thin ~5-atomic-layer film of liquid helium at three different temperatures: T=0.4K, 0.98K and 1.3K. 
Similar and even these data were partially published previously in Refs. \cite{Godfrin1992,GodfrinJLTP1992,Clements1996}, but here we present them at different temperatures, in a better quality and in color for greater visibility and resolution \cite{GodfrinPrivate}. 

The experimental setup and method were described previously in
detail.\cite{Godfrin1992,GodfrinJLTP1992,Clements1996} Helium was adsorbed onto a substrate
of exfoliated graphite. The He film of the thickness of approximately 5
atomic layers has been used, because for thicker films the contribution of
surface excitations is too weak as compared to the dominant contribution
from the bulk excitations (phonons). Thinner films also have drawbacks for
the study of surfons. First, the 2-3 atomic layers adjacent to the substrate
are solid and their structure differs considerably from that in the bulk He.
Second, due to the dimensional quantization along z-axis, the bulk
excitations in too thin films may also contain the energy gap $\sim 1K$ and
resemple the surfons. The inelastic neutron-scattering experiments were
performed at the ILL on the time-of-flight spectrometer IN6 with an incident
wavelength of 5.12\AA . The detectors were located in an angular range,
corresponding to momentum transfers between $0.254$\AA $^{-1}$ and $2.046$%
\AA $^{-1}$ for elastically scattered neutrons. The energy resolution was $%
\sim 0.6-0.7$K and depended only slightly on the momentum transfer. In Ref. 
\cite{Godfrin1992,GodfrinJLTP1992,Clements1996} the authors studied mainly the ripplon and
phonon spectra in several He films of various thickness and in Ref. 
\cite{Clements1996} in the shorter interval $%
q_{||}>0.4\mathring{A}^{-1}$ of in-plane wave-vector. In
this paper we present these experimental data in a full available wave-vector interval 
$q_{||}>0.25\mathring{A}^{-1}$\ and concentrate on the momentum-energy
region corresponding to the expected dispersion of surfons. In addition, we
provide partially unpublished data \cite{GodfrinPrivate} at three different temperatures of liquid He, namely, $T=0.4K$, $%
0.98K$ and $1.3K$, while in Refs. \cite{Godfrin1992,GodfrinJLTP1992,Clements1996} only the
data at $T=0.65K$ are given.

%In Figs. \ref{FigT098}-\ref{FigT13} we present the intensity plots of inelastic neutron scattering as a function of the energy $\hbar \omega $ and in-plane wave-vector $q_{\parallel }$ of induced excitations at two different temperatures $T=0.4K$ and $T=1.3K$. These data start at $q_{\parallel }=0.25\mathring{A}^{-1}$, covering a larger low-momentum interval than those given previously in Refs. \cite{Godfrin1992,Clements1996}. 
%At $q_{\parallel }>0.4\mathring{A}^{-1}$ The phonon and ripplon modes agree with those in Refs. \cite{Godfrin1992,Clements1996}. 

The intensity of neutron scattering in Figs. \ref{FigT098}-\ref{FigT13} 
as a function of the energy $\hbar \omega $ and in-plane wave-vector 
$q_{\parallel }$ of induced excitations is given by color
(brightness in greyscale) as shown on the left panel of Fig. \ref{FigT098}.
The bright areas form thick lines in the $q_{\parallel }-\omega $
coordinates, which give the dispersion relation of excitations. In all Figs. %
\ref{FigT098}-\ref{FigT13} one can easily distinguish the ripplon branch
(lowest curve) and the phonon branch (upper curve), marked by white thin
curves. The phonon and ripplon modes agree with those in Refs. \cite{Godfrin1992,GodfrinJLTP1992,Clements1996}.
 The phonon branch has a roton minimum at $q_{\parallel }\approx 2%
\mathring{A}^{-1}$, which gives a strong intensity maximum of inelastic
neutron scattering. In addition to these two well-know excitation branches,
on each of Figs. \ref{FigT098}-\ref{FigT13} one can distinguish another
curve of intensity maxima, located at $0.25\mathring{A}^{-1}<q_{\parallel
}<1.5\mathring{A}^{-1}$ between the phonon and ripplon branches and
approximately coinciding with the white dashed line of expected surfon dispersion.
The intensity of neutron scattering (brightness) of this curve is beyond the
error-bar \cite{GodfrinPrivate} and at small wave-vector is even stronger 
than that of ripplons on
all three Figs. \ref{FigT098}-\ref{FigT13}. This intermediate dispersion
curve gives a gapped excitation and, possibly, originates from surfons,
because the white dashed lines in Figs. \ref{FigT098}-\ref{FigT13} show the
expected surfon dispersion given by Eq. (\ref{Disp}) with the activation
energy $\Delta \approx 4.5K$. The corresponding effective mass of
surfon branch coincides with the mass of a free He atom, $M^{\ast }=M_{4}^{0}$%
, because the inelastic neutron scattering is a process of short time $\sim
\hbar /\varepsilon <\hbar /\Delta $, while the "dressing" of surfons by the
formation of the ripplonic polaron (dimple), leading to the increase of
surfon effective mass,\cite{SurStates2011} requires longer time. Note that some traces of this
additional branch of surface excitations are seen already in Fig. 1 of Ref. 
\cite{Godfrin1992}, but in the Figs. \ref{FigT098}-\ref{FigT13} of current paper
this branch is clearer. Thus, the available experimental data on inelastic neutron scattering by thin He films support the existence of surfons \cite{SurStates,SurStates2011,WeJETPLett2008,MobilityLett} as an additional type of surface excitations.

\section{Comparison of various theoretical calculations of surface excitations in helium and
discussion}

The microscopic numerical calculations also propose several types of surface
excitations in addition to ripplons.\cite%
{Krotscheck1985II,Krotscheck1985Substrate,KrotscheckTymczak,ClementsKrotscheckPRB1996,ClementsKrotscheckJLTP1997,Gernoth1992,Gernoth1994}
These numerical calculations apply the correlated basis function (CBF)
method \cite{Mahan} for inhomogeneous liquid, using some additional
approximations. These CBF calculations are based on the Feenberg wave
function with only pair correlations and performed in the hypernetted-chain
approximation. These calculations also assume that only one-body component
of the Feenberg function is affected by excitations and by external
perturbations.\cite{Krotscheck1985II} This assumption of static two-body
correlations limits the regime of validity of this theory to wavelengths
longer than the average distance between two particles. It also may restrict
the theory to small deviations from the equilibrium (ground-state) density
of liquid He. The backflow effects \cite{FenbergBook1969} are also ignored in
these numerical calculations. Therefore, the obtained theoretical excitation
energies calculated at large wave numbers $q_{\parallel }$ are substantially
higher than the experimental results.\cite{Gernoth1994} Since the applied
Jastrow variational treatment of the bulk liquid does not produce a
self-bound system at saturation He density, an external potential is
introduced phenomenologically in these numerical calculations to stabilize
the surface.\cite{Gernoth1992,Gernoth1994} The strength of this additional
phenomenological potential is adjusted so that the calculated chemical
potential matches the experimental saturation value.\cite{Gernoth1994}
Finally, the standard numerical computations assume that surface excitations
do not violate the translational symmetry along the surface, which may not
describe the case of a single surfon with zero in-plane momentum.
Nevertheless, the comparison of the surface excitations proposed by these
approximate numerical calculations with the surfons proposed
semi-phenomenologically \cite{SurStates,SurStates2011} is quite useful.

In Refs. \cite%
{Krotscheck1985II,Krotscheck1985Substrate,KrotscheckTymczak,ClementsKrotscheckJLTP1997}
the ground state and excitations in thin He film of several atomic layers on
a substrate of various materials were investigated, and several types of
excitations in He films with a non-zero gap were found. The analysis of the
particle currents and transition densities, in addition to the dispersion
relation of these excitations, allowed to describe their nature:\cite%
{ClementsKrotscheckJLTP1997} they were attributed to the ripplon mode on the
He-substrate interface and to the so-called \textquotedblright breathing
mode\textquotedblright . The latter describes the standing wave in the $z$%
-direction perpendicular to the film,\cite{ClementsKrotscheckJLTP1997}
similar to the volume phonon, which may propagate along the film. The energy
gap of this \textquotedblright breathing mode\textquotedblright\ reduces
with the increase of the film thickness. Thus, no one of the excitations
found in Refs. \cite%
{Krotscheck1985II,Krotscheck1985Substrate,KrotscheckTymczak,ClementsKrotscheckJLTP1997}
can be attributed to surfons. The reason is that these calculations \cite%
{Krotscheck1985II,Krotscheck1985Substrate,KrotscheckTymczak,ClementsKrotscheckJLTP1997}
include only excitations inside the liquid, neglecting He vapor and the
states above liquid He. This restriction was eliminated in Refs. \cite%
{Gernoth1992,Gernoth1994}, where a free boundary between a deep liquid He
and saturated vapor was investigated by the similar numerical microscopic
CBF approach, and notably different results were obtained. This limit of
deep liquid He, filling a half-space instead of atomically thin film, is
closer to the model of semi-phenomenological description of surfons in Ref. 
\cite{SurStates2011}. Again several gapped surface excitations were obtained
in deep liquid He \cite{Gernoth1992,Gernoth1994}, but the structure of these
excitations is completely different from those in thin films \cite%
{Krotscheck1985II,Krotscheck1985Substrate,KrotscheckTymczak,ClementsKrotscheckJLTP1997}%
. First, no signature of \textquotedblright breathing
mode\textquotedblright\ was found at the free surface of bulk He,\cite%
{Gernoth1992,Gernoth1994} which is natural because this mode was found to
spread along the whole He film thickness \cite{ClementsKrotscheckJLTP1997},
being rather the bulk excitation. Nevertheless, two new types of gapped
surface excitations in deep He were found.\cite{Gernoth1992} The first type
has a large excitation energy above $\Delta _{1}\approx 18.5K$ and is
interpreted as a bound roton trapped in the interface region.\cite%
{Gernoth1992} The wave function of this excitation is mostly inside liquid
He (see Fig. 2 of Ref. \cite{Gernoth1992}), so it cannot be interpreted as
the surfon. 

The second type of gapped surface excitations, found in Ref. \cite{Gernoth1992} 
and called resonance interface state (RIS), has the structure and
properties very similar to those of surfons. First, RIS correspond to a peak
of He density just above the liquid surface, as shown in Fig. 6 of Ref. \cite%
{Gernoth1992}. This is very similar to the wave function of surfons, shown
in Fig. 1 of Ref. \cite{SurfonEvaporation}. Second, RIS in-plane dispersion
is very similar to that of surfons (see Figs. 7 and 9 in Ref. \cite%
{Gernoth1992}): their energy gap $\Delta _{2}\approx 1.6K$, and at in-plane
momentum $k_{\parallel }<1.5\mathring{A}^{-1}$ they have almost a quadratic
dispersion in Eq. (\ref{Disp}) with effective mass $M^{\ast }$ close to the
atomic He mass $M_{4}^{0}=6.7\cdot 10^{-24}$g. Third, similar to surfons,
these surface excitations are interpreted as the He vapor atoms with wave
function having large peak just above liquid surface \cite{Gernoth1992},
forming a bound state at the surface at zero temperature. At finite
temperature these atoms in the bound surface states are quasi-stationary,
i.e. they have small finite probability to become delocalized, similar to
surfon evaporation at finite temperature studied in Ref. \cite%
{SurfonEvaporation}. Thus we suggest that the resonance interface states,
obtained numerically in Ref. \cite{Gernoth1992}, and the surfons, proposed
in Refs. \cite{SurStates,SurStates2011}, describe the same type of surface
excitations with two different approximate approaches.\footnote[1]{The second 
	surface excitation mode, proposed in Ref. \cite{Gernoth1994} for deep He, 
	probably, corresponds to the superposition of surfon and surface phonon mode, 
	as it has two maxima of the square of He wave function, one above and 
	one below the He surface.} 

Therefore, the experimental investigation of the dispersion law of surface
excitations, provided by inelastic neutron scattering on He films, is very
helpful for detecting surfons and studing their properties. These properties
may somewhat differ from those predicted by the semi-phenomenological
approach of Refs. \cite{SurStates,SurStates2011} or approximate numerical
calculations of Ref. \cite{Gernoth1992}. The numerical calculations of the
dynamics of one He atom approaching the surface and interacting with nearest
atoms from the liquid could additionally prove the existence of surfons and
even estimate their lifetime. Taking into account the important role of
surfon excitations in the physical properties of the surface of liquid
helium and, possibly, of other cryogenic liquids, further numerical
calculations on this problem are highly need.

The observed additional branch of intensity maxima, giving the in-plane
dispersion of surface excitations and approximately coinciding with the
dashed line in Figs. \ref{FigT098}-\ref{FigT13} of possible surfon spectrum,
give a strong support of the existence of surfons and suggest their
dispersion law. Alternatively, this additional branch could be due to the
\textquotedblright breathing mode\textquotedblright ,\ obtained in Refs. 
\cite%
{Krotscheck1985II,Krotscheck1985Substrate,KrotscheckTymczak,ClementsKrotscheckJLTP1997}
for thin He films. This breathing mode has a different in-plane dispersion,
closer to linear rather than quadratic as for surfon. In addition, for thick
films there should be several such modes, corresponding to different quantum
numbers of dimensional quantization along the z-axis. The monitoring of the
evolution of the activation energy of this mode with the change of He film
thickness could elucidate the nature of this excitation and completely rule
out (or confirm) its breathing-mode origin, but this, probably, requires
experimental data with higher energy resolution. 
%However, we don't have the data for different film thickness (???).

\medskip

To summarize, we analyze the experimental data on inelastic neutron scattering by thin $\sim 5$-atomic-layer film of liquid helium at three different temperatures. The scattering intensity plot, shown in Figs. \ref{FigT098}-\ref{FigT13}, suggests a new type of gapped surface excitations with activation energy $\sim 4.5$K and dispersion similar to that expected for surfons, proposed and investigated semi-phenomenologically in Refs. \cite{SurStates,SurStates2011,WeJETPLett2008,MobilityLett}. Surface excitations with very similar structure and properties were also obtained by numerical calculations and called resonance interface states.\cite{Gernoth1992} Before there were only indirect experimental substantiations of surfons, based on temperature dependence of surface tension coefficient \cite{SurStates,SurStates2011} and on interaction of surfons with surface electrons \cite{WeJETPLett2008,MobilityLett}. The shown data on inelastic neutron scattering, probably, provide the first direct observation of surfons. However, further experimental and theoretical study is need for the undoubted confirmation of surfons as surface excitations and for the quantitative analysis of their properties.

\medskip

The authors thank H. Godfrin for providing the partially unpublished experimental data \cite{GodfrinPrivate} and E. Krotscheck for useful discussions. The work was
supported by the program 0033-2018-0001 \textquotedblleft Condensed Matter
Physics\textquotedblright\ by the FASO of Russia. A.D.G. thanks RFBR grant
\#16-02-00522.


\begin{thebibliography}{99}
		
\bibitem{GodfrinPrivate} H.J. Lauter and H. Godfrin, private communication.

\bibitem{Godfrin1992} H. J. Lauter, H. Godfrin, V. L. P. Frank, and P.
Leiderer, Phys. Rev. Lett. \textbf{68}, 2484 (1992).

\bibitem{GodfrinJLTP1992} H. J. Lauter, H. Godfrin, and P.
Leiderer, J. Low Temp. Phys. \textbf{87}, 425 (1992).

\bibitem{Clements1996} B. E. Clements, H. Godfrin, E. Krotscheck, H. J.
Lauter, P. Leiderer, V. Passiouk, and C. J. Tymczak, Phys. Rev. B \textbf{53}%
, 12242 (1996).

\bibitem{SurStates} A.M. Dyugaev, P.D. Grigoriev, JETP Lett. \textbf{78},
466 (2003) [Pisma v ZhETF \textbf{78}, 935 (2003)].

\bibitem{SurStates2011} A.D. Grigoriev, P.D. Grigoriev, A.M. Dyugaev, J. Low
Temp. Phys. \textbf{163}, 131 (2011); arXiv:0905.2306.

\bibitem{MobilityLett} P.D. Grigoriev, A.M. Dyugaev, E.V. Lebedeva, JETP 
\textbf{106}(2), 316 (2008).

\bibitem{WeJETPLett2008} P. D. Grigor'ev, A. M. Dyugaev and E. V. Lebedeva,
JETP Letters \textbf{87}, 106 (2008) [Pisma v ZhETF \textbf{87}, 114 (2008)].

\bibitem{SurfTensionBook} J.S.Rowlinson and B. Widom, \textit{Molecular
Theory of cappilarity}, Dover Publications, Mineola NY, 2002.

\bibitem{EdwardsReview1978} D.O. Edwards and W.F. Saam, Chapter 4 in \textit{%
The free surface of liquid helium}, Ed. by D.F. Brewer, Progress in Low
Temperature Physics (series), North-Holland Publishing Company, 1978

\bibitem{Edelman} V.S. Edel'man, Sov. Phys. - Uspehi \textbf{130}, 676
(1980).

\bibitem{Shikin} V.B. Shikin and Yu.P. Monarkha, \textit{Two-Dimensional
Charged Systems in Helium } (in Russian), Nauka, Moscow (1989).

\bibitem{Monarkha} Y. Monarkha, K. Kono, \textit{Two-Dimensional Coulomb
Liquids and Solids}, Springer Verlag, 2004.

\bibitem{FenbergBook1969} E. Feenberg, \textit{Theory of Quantum Fluids}
(Academic, New York, 1969).

\bibitem{Mahan} G. Mahan, Many-Particle Physics, 2nd ed. (Plenum Press, New
York, 1990), Ch. 10.

\bibitem{Krotscheck1985II} E. Krotscheck, Phys. Rev. B \textbf{31}, 4258
(1985).

\bibitem{Krotscheck1985Substrate} E. Krotscheck, Phys. Rev. B \textbf{32},
5713 (1985).

\bibitem{KrotscheckTymczak} E. Krotscheck and C. J. Tymczak, Phys. Rev. B 
\textbf{45}, 217 (1992).

\bibitem{ClementsKrotscheckPRB1996} B. E. Clements, E. Krotscheck, and 
C. J. Tymczak, Phys. Rev. B \textbf{53}, 12253 (1996).

\bibitem{Gernoth1994} K.A. Gernoth, J.W. Clark, G. Senger and M.L. Ristig,
Phys. Rev. B \textbf{49}, 15836 (1994).

\bibitem{Gernoth1992} K. A. Gernoth and M. L. Ristig, Phys. Rev. B \textbf{45%
}, 2969 (1992).

\bibitem{ClementsKrotscheckJLTP1997} B. E. Clements, E. Krotscheck, C. J. 
Tymczak, J. Low Temp. Phys. \textbf{107}, 387 (1997).

\bibitem{Lauter} H.-J. Lauter, in \textit{Excitations in Two-Dimensional and
	Three Dimensional -Quantum Fluids}, NATO ASI series, edited by A. F. G.
Wyatt aud H. J. Lauter (Plenum, New York, 1991).

\bibitem{Pricaupenko} L. Pricaupenko and J. Treiner, J. Low Temp.Phys. 
\textbf{101}, 809 (1995).

\bibitem{CampbellPRB1997} C. E. Campbell, B. E. Clements, E. Krotscheck, and M. Saarela, Phys. Rev. B \textbf{55}, 3769 (1997).

\bibitem{LL6} L. D. Landau and E. M. Lifshitz, Course of Theoretical
Physics, Vol. 6: Fluid Mechanics, 2nd ed. Butterworth-Heinemann, 1987.

\bibitem{AndreevLevels} A.F. Andreev, JETP \textbf{23}, 939 (1966) [Zh. Exp.
Teor. Fiz. \textbf{50}, 1415 (1966)].

\bibitem{Hydrogen1} C.G. Paine and G.M. Seidel, Phys. Rev. B \textbf{46},
1043 (1992).

\bibitem{ExpMobil} K.Shirahama, S. Ito, H.Suto and K. Kono, J. Low Temp.
Phys. \textbf{101}, 439 (1995).

\bibitem{Saitoh} M. Saitoh, J. Phys. Soc. Japan \textbf{42}, 201 (1977).

\bibitem{HeDataBook} Russell J. Donnelly and Carlo F. Barenghi, \textit{The
	Observed Properties of Liquid Helium at the Saturated Vapor Pressure,}
Journal of Physical and Chemical Reference Data 27, 1217 (1998).

\bibitem{SurfonEvaporation} A.D. Grigoriev, P.D. Grigoriev, A.M. Dyugaev,
A.F. Krutov, Low Temp. Phys. \textbf{38}, 1005 (2012).

\bibitem{KrotscheckPrivate} E. Krotscheck, private communication.

\bibitem{TuckerJLTP1995} M. A. H. Tucker and A. F. G. Wyatt, Journal of Low Temp. Phys. (Springer)  \textbf{100}, 105 (1995).

\end{thebibliography}
\end{document}